# How learning environment predicts male and female students' physics motivational beliefs in introductory physics courses

Yangqiuting Li, Kyle Whitcomb, and Chandralekha Singh

*Department of Physics and Astronomy, University of Pittsburgh, Pittsburgh, PA 15260*

In this study, we adapt prior identity framework to investigate the effect of learning environment (including perceived recognition, peer interaction and sense of belonging) on students' physics self-efficacy, interest and identity by controlling for their self-efficacy and interest at the beginning of a calculus-based introductory physics course. We surveyed 1203 students, 35% of whom were women. We found that female students' physics self-efficacy and interest were lower than male students' at the beginning of the course, and the gender gaps in these motivational constructs became even larger by the end of the course. Analysis revealed that the decrease in students' physics self-efficacy and interest were mediated by the learning environment and ultimately affected students' physics identity. Our model shows that perceived recognition played a major role in explaining students' physics identity, and students' sense of belonging in physics played an important role in explaining the change in students' physics self-efficacy.

## I. INTRODUCTION

Some prior studies have focused on underrepresented groups such as women in science, technology, engineering, and mathematics (STEM) courses, majors and careers [1-12]. Prior research suggests that individuals' course enrollment, degree attainment and achievement in STEM can be influenced by their motivational beliefs such as self-efficacy, interest and identity in that domain [2,4,6,13-19]. For students from underrepresented groups, these motivational beliefs are shaped, e.g., by negative societal stereotypes and biases about who belongs in STEM and can excel in STEM as well as lack of role models and encouragement from others, and they can lead to withdrawal from STEM courses, majors or careers [20-26]. Hence, investigating students' motivational characteristics is critical to understanding and addressing diversity, equity, and inclusion issues in the STEM disciplines.

In physics, researchers have found that self-efficacy is an important motivational characteristic of students in order to excel [4-7]. Self-efficacy is the belief in one's capability to be successful in a particular task, course, or subject area [27,28], and it has been shown to influence students' engagement and performance in a given domain [15,17,18,29]. Another important motivational characteristic of students is interest, which is defined by positive emotions accompanied by curiosity and engagement in a particular discipline [30,31]. According to Eccles's expectancy-value theory [32,33], interest is paired well with self-efficacy as connected constructs that predict students' academic outcomes and career aspirations. In addition, science or physics identity is another motivational characteristic that can influence students' career decisions and outcome expectations [1-3,34-38]. Students' science identity, e.g., is related to whether they see themselves as a science person [1-3,34,35,38].

Prior studies show that female students consistently report lower self-efficacy than male students in many STEM courses [4-6]. In addition, female students are less likely to see themselves as a physics person than male students [2,39]. Therefore, investigating the factors that influence physics motivational beliefs in introductory physics courses taken by physical science and engineering majors can play an important role in understanding women's underrepresentation in those disciplines. Here we describe a study focusing on how the perception of the learning environment predicts students' physics self-efficacy, interest and identity in a calculus-based introductory physics course. Our findings can be useful for developing an inclusive and equitable learning environment.

## II. BACKGROUND AND FRAMEWORK

The well-known science identity framework by Carlone and Johnson [1] includes three dimensions: competence ("I think I can"), performance ("I am able to do"), and recognition ("I am recognized by others"). Hazari et al. added interest to this framework and pointed out that the relation between gender and physics identity was mediated by interest, competency belief, and perceived recognition [40,41]. These studies reveal that individuals' internal identity in science is impacted by their perceived recognition from others.

Similarly, students' self-efficacy and interest have also been found to be influenced by their interaction with others [28,31]. According to Bandura's social cognitive theory, one factor that contributes to the development of self-efficacy is social persuasion experiences. In Hidi and Renninger's four stages model of interest development [31,42,43], people's interest in a discipline is triggered and maintained by external factors first, but then it becomes an individual interest and finally becomes a well-developed interest. In prior work [44], students' perceived recognition is not only the strongest predictor of identity, it also predicts self-efficacy and interest.

In addition to perceived recognition, some studies have shown that students' interactions with peers and students' sense of belonging are also important parts of their perceptions of the learning environment [45-51]. For example, if students have a higher sense of belonging, they may approach others in the academic environment more often and with more positive attitudes, building better interactions with others and reporting higher perceived recognition from others [52]. However, there are few quantitative studies about the effect of learning environment on students' physics motivational beliefs and the roles played by each component of the learning environment. Thus, to better understand how the learning environment influences student outcomes and how to foster an inclusive and equitable learning environment, further study is needed.

Inspired by the framework that learning environment can play a key role in shaping female and male students' motivational beliefs, we investigated how the perception of learning environment (including perceived recognition, peer interaction and belonging) predicts students' physics self-efficacy, interest and identity at the end of the course (post) by controlling for students' self-efficacy and interest at the beginning (pre) of a calculus-based introductory physics course. The learning environment in this work here is not limited to the classroom environment: it also includes scenarios outside the class. For example, students may work together on their homework after class, and they may also ask for help during instructors' office hours. As shown in Fig. 1, the nine constructs can be divided into three groups: what we control for, perceptions of learning environment, and motivational outcomes. Students' gender, pre-self-efficacy (pre SE) and pre-interest measured at the beginning of the physics course are constructs that we control for, which are related to students' beliefs about physics based on their prior experiences. Outcomes include students' post-self-efficacy (post SE), post-interest and identity measured at the end of the course. Perceived recognition (recog), peer interaction (int) and sense of belonging (bel) constitute the learning environment. They are also measured at the end of the course, because only after the course can students have a good estimate of these based on their real experiences about interacting with peers, TAs and instructors.

In this study we first estimated gender differences in student's pre-self-efficacy and interest. Then, we studied how students' self-efficacy and interest changed from pre to post and how much of those change can be explained by the learning environment. Also, to better understand the roles played by each of the three learning environment factors, we first used path analysis, which only includes perceived recognition, and investigated how much of students' physics self-efficacy, interest and identity were explained by the model. Then we added peer interaction and belonging into this model one by one to investigate how much extra variance in student outcomes is explained by these learning environment factors. If variance in one environmental factor is correlated with student motivational outcomes' variance not covered by the other two learning environment factors, then adding that factor will explain additional variance in student outcomes.

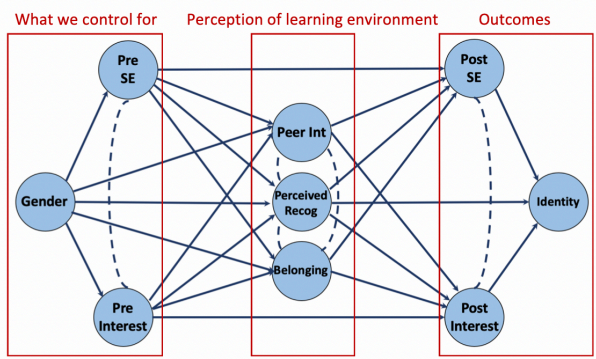

FIG. 1. Schematic representation of the framework. In the path analysis, from left to right, all possible regression paths were considered, but only some paths are showed here for clarity.

### III. RESEARCH QUESTIONS

Our research questions regarding student outcomes in the first calculus-based introductory physics course taken by engineering and physical science majors in their first year of college studies at a large state-related research university in the United States are as follows:

**RQ1.** Are there gender differences in students' physics motivational characteristics and do they change from the beginning to the end of the course (i.e., from pre to post)?

**RQ2.** How do the components of the learning environment (including perceived recognition, peer interaction and belonging) predict students' physics identity as well as post-self-efficacy and post-interest controlling for pre-self-efficacy and pre-interest?

**RQ3.** Does the strength of the relationships given by the standardized regression coefficients between any two constructs in the models differ for women and men? If not, how does gender predict each construct in the model?

### IV. METHODOLOGY

In this study, we collected pre and post motivational survey data from students who took the introductory calculus-based physics 1 course in two consecutive fall semesters. The course was taught in a traditional, lecture-based format and was taken mostly by students majoring in engineering and other physical sciences. The paper surveys were handed out and collected by TAs in the first and last recitation class of a semester. We named the data collected at the beginning of the semester as pre-data and that collected at the end of the semester as post-data. Finally, we combined the two semesters' data and put them into the two categories, pre and post. The demographic data of students—such as gender—were provided by the university. Students' names and IDs were de-identified by an honest broker who provided each student with a unique new ID (which connected students' survey responses and their demographic information). Thus, researchers could analyze students' data without having access to students' identifying information. There were 1203 students who participated in the study including both year (427 female students and 776 male students). We recognize that gender identity is not binary. However, because students' gender information was collected by the university which offered binary options, we did the analysis with the binary gender data available.

In this study, we considered six constructs—physics self-efficacy, interest, peer interaction, perceived recognition, belonging, and identity. The survey questions were adapted from the prior motivational surveys [40,41,53-56] and were re-validated in our prior work [6,44]. The validation and refinement of the survey involved use of one-on-one student interviews with both introductory and advanced students, exploratory and confirmatory factor analyses (EFA and CFA), Pearson correlation between factors and Cronbach alpha [57].

In our survey, we had four items for self-efficacy (Cronbach alpha = 0.69 for pre-self-efficacy, Cronbach alpha = 0.8 for post-self-efficacy [57]). Students had four options on a Likert scale for each item which corresponded to 1 to 4 points. We also had four items for interest; each item involved a four-point Likert scale (Cronbach alpha = 0.75 for pre-interest, Cronbach alpha = 0.82 for post-interest). Physics identity corresponds to students' belief about whether they see themselves as a physics person [3]. Students could choose from strongly disagree, disagree, agree, and strongly agree and they corresponded to 1 to 4 points [58], respectively.

In addition, perceived recognition, peer interaction and belonging are the other constructs for learning environment. Unlike self-efficacy, interest and identity, these three constructs are directly related to students' experience in the course. Perceived recognition included three items which represent whether a student thinks other people see them as a physics person [2,3,34] (Cronbach alpha = 0.86). Peer interaction includes four items and represents whether students have a productive experience when working with peers (Cronbach alpha = 0.91). Belonging is about students' feelings of whether they thought they belonged in the physics class [48]. Consistent with the prior survey from which these items were taken, students had five options for all five belonging items: Not at all true, a little true, somewhat true, mostly true and completely true (Cronbach alpha = 0.86).

First, we calculated the mean score for each construct for each student. Then we used a *t*-test [59,60] to compare students' pre- and post-scores and to compare responses for female and male students. Finally, we used Structural

Equation Modeling (SEM) [61] to analyze predictive relationships between constructs using our survey data. The SEM includes two parts: CFA and path analysis (see Fig. 1).

To validate the items on our survey, we performed the CFA for each construct. The model fit is good if the fit parameters are above threshold. In CFA, Comparative Fit Index (CFI) > 0.9, Tucker-Lewis Index (TLI) > 0.9, Root Mean Square Error of Approximation (RMSEA) < 0.08 and Standardized Root Mean Square Residual (SRMR) < 0.08 are considered as acceptable and RMSEA < 0.06 and SRMR < 0.06 are considered as a good fit [62]. In our study, CFI = 0.943, TLI = 0.934, RMSEA = 0.05 and SRMR = 0.041, which represents a good fit. Thus, there is additional quantitative support for dividing the constructs as proposed.

To analyze the relations among the constructs, we performed the full SEM. Apart from CFA, SEM gives regression coefficients $\beta$ for paths between each pair of constructs and the value of each $\beta$ is a measure of the strength of that relationship. Compared with a multiple regression model, the advantage of SEM is that we can estimate all the regression links for multiple outcomes and factor loadings for items through CFA simultaneously [62]. Before performing the gender mediation models, we first tested the gender moderation relations between each pair of constructs using multi-group SEM. Results showed that in all of our models, strong measurement invariance holds and there is no difference in any regression coefficients by gender, which allowed us to perform the gender mediation analysis using SEM [62]. In this study, we first considered a model with perceived recognition as the only learning environment to see how students' physics self-efficacy, interest and identity were predicted by it. Then, we added peer interaction or belonging as additional constructs in the learning environment. Finally, our model included all of the three learning environment constructs (see Fig. 1). We analyzed the variance in each construct (factor or latent variable) denoting student outcome explained by each model to understand the unique role played by each learning environment and to determine if all three learning environment components are productive.

## V. RESULTS AND DISCUSSION

As shown in Table I, female students had significantly lower average interest and self-efficacy scores than male students in both the pre- and post-survey and these gender gaps increased by the end of the semester. The effect size given by Cohen's $d$ [60] for gender difference in physics interest increased from 0.54 to 0.60, and the effect size of gender difference in self-efficacy increased from the 0.32 to the 0.53. In addition, even though students' self-efficacy dropped generally from pre to post (see Table I), female students' self-efficacy dropped (effect size $d = 0.52$) more than male students' (effect size $d = 0.27$). Similarly, Table I shows that female students' interest in physics dropped (effect size $d = 0.30$) more than male students' (effect size $d = 0.19$). Table II shows the average scores on other constructs (perception of peer interaction, perceived recognition, belonging and identity) in the post-survey. Female students had significantly lower average scores in all of these constructs with the effect sizes for all of them in the medium range.

**Table I.** Descriptive statistics for pre- and post-self-efficacy and interest in physics by gender.

| Gender | Self-efficacy (1-4) | | Statistics | |
|---|---|---|---|---|
| | Pre- | Post- | $p$ value | Cohen's $d$ |
| Male | 3.12 | 2.98 | <0.001 | 0.27 |
| Female | 2.96 | 2.68 | <0.001 | 0.52 |
| $p$ value | <0.001 | <0.001 | | |
| Cohen's $d$ | 0.32 | 0.53 | | |
| Gender | Interest (1-4) | | Statistics | |
| | Pre- | Post- | $p$ value | Cohen's $d$ |
| Male | 3.19 | 3.08 | <0.001 | 0.19 |
| Female | 2.89 | 2.70 | <0.001 | 0.30 |
| $p$ value | <0.001 | <0.001 | | |
| Cohen's $d$ | 0.54 | 0.60 | | |

**Table II.** Descriptive statistics for perceived recognition, peer interaction, belonging and identity in physics by gender.

| Gender | Perception of learning environment | | | Identity (1-4) |
|---|---|---|---|---|
| | Peer Int (1-4) | Perceived Recog (1-4) | Belonging (1-5) | |
| Male | 2.97 | 2.60 | 3.73 | 2.63 |
| Female | 2.68 | 2.24 | 3.33 | 2.17 |
| $p$ value | <0.001 | <0.001 | <0.001 | <0.001 |
| Cohen's $d$ | 0.44 | 0.49 | 0.46 | 0.56 |

Finally, for determining the predictive relationships between constructs using SEM, because many studies have shown that perceived recognition is a strong predictor of students' motivational beliefs [44,63-65], all of the models we tested include perceived recognition as one of the learning environment constructs. First, perceived recognition was the only learning environment construct we included in the model. Then we added peer interaction or belonging to the learning environment one by one to analyze how each helped to predict students' post-self-efficacy and interest. Finally, we included all of the three constructs in our model and studied how these constructs mediated the outcomes together and what role was played by each of them. The results of the last SEM model are presented visually in Figure 2. The model fit indices of all of the models show good model fits to the data (for example, for the last model, CFI = 0.941 (>0.90), TLI = 0.932 (>0.90), RMSEA = 0.049 (<0.08) and SRMR = 0.042 (<0.08)).

We find that all three learning environment constructs predict students' self-efficacy and interest at the end of the course even after controlling for students' pre-self-efficacy and pre-interest. According to Figure 1, belonging is the largest predictor of post-self-efficacy. Although identity is predicted by self-efficacy, interest and perceived recognition, perceived recognition, which is the only environmental factor of these three, is the largest predictor.

Another interesting finding is that as we added more constructs to the learning environment, the strength of the direct paths from pre to post decreased since the learning environment factors mediated learning. For example, the direct effect of pre-self-efficacy on post-self-efficacy is 0.35 in the model which only has perceived recognition, while this effect decreased to 0.29 when we added peer interaction, and decreased to 0.23 in the last model which includes all three learning environment factors. These results indicate that the learning environment is mediating the effect of students' pre-self-efficacy and pre-interest on their outcomes.

**Table III.** Coefficient of determination ($R^2$) for various constructs in different models showing how different combinations of the perception of learning environment predict various student outcomes. All $R^2$ values are significant with p < 0.001. Recog, peer, bel are abbreviations for perceived recognition, peer interaction and sense of belonging.

| Construct | SEM Models with Seven Different Learning Environment Factors | | | | | | |
|---|---|---|---|---|---|---|---|
| | Recog | Peer | Bel | Peer+Recog | Peer+Bel | Recog+Bel | Peer+Recog+Bel |
| Post-SE | 0.56 | 0.61 | 0.75 | 0.68 | 0.77 | 0.77 | 0.79 |
| Post-Interest | 0.79 | 0.77 | 0.78 | 0.80 | 0.78 | 0.80 | 0.80 |
| Identity | 0.74 | 0.61 | 0.61 | 0.74 | 0.61 | 0.75 | 0.75 |

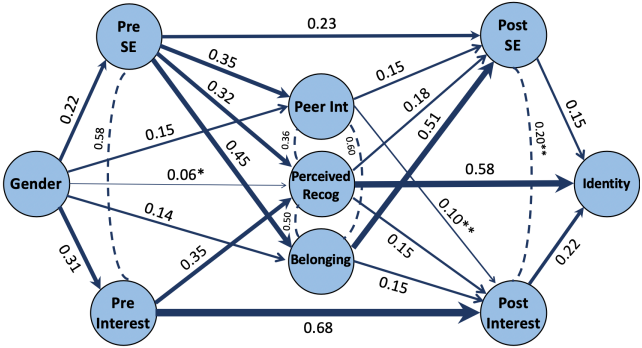

FIG. 2. Schematic diagram of the path analysis part of the SEM with gender mediation. The regression line thickness corresponds to the magnitude of $\beta$ (standardized regression coefficient) with $0.01 < p < 0.05$ indicated by * and $0.001 < p < 0.01$ indicated by **. Other regression lines show relations with $p < 0.001$.

Although there are large gender differences in students' average pre- and post- self-efficacy and interest, Fig. 2 shows gender mediation and clarifies that gender only directly predicts pre-self-efficacy, pre-interest, and the perception of learning environment. Thus, Fig. 2 reveals that the gender differences in students' post-self-efficacy, post- interest, and post-identity shown in Table I were mediated by the different components of the learning environment.

To further understand the role played by each learning environment construct, Table III shows the coefficient of determination $R^2$ (fraction of variance explained) for each of the three student outcomes for seven different SEM models with different combinations of components of the learning environment. As shown in Table III, $R^2$ values for post-interest are around 0.79 in all of the models. This means the models which include any of the three learning environment constructs can explain 79% of the variance of post-interest. However, there is 75% of the variance of post-self-efficacy explained by the model which only includes belonging, which is larger than that explained by the other two single construct models and is very close to that explained by the model including all of the three constructs. Similarly, the model which only includes perceived recognition explains 74% of the variance of identity, and adding peer interaction and belonging does not help explain the variance in identity further. Table III shows that both belonging and perceived recognition play unique roles in the learning environment in explaining outcomes. However, peer interaction co-varies with belonging and perceived recognition and uniquely explains very small percentages of the variance in the outcomes in Table III. The co-variation suggests a possibility that students' sense of belonging and perceived recognition can be shaped by helping students interact meaningfully with peers (which in turn can improve their learning outcomes). Thus, we believe the model including all of the three learning environment constructs is productive.

## VI. SUMMARY AND CONCLUSION

Due to reasons such as societal stereotypes and biases about who belongs in physics and can excel in it, women continue to be disadvantaged in physics courses and major as well as in related disciplines such as engineering. We found significant gender differences favoring male students in all motivational constructs in our models for students in a calculus-based introductory physics course. In addition, we found that both male and female students' self-efficacy and interest dropped from pre to post and female students' dropped even more than male students'. We went beyond expectancy-value theory and included in our SEM model components of the learning environments. In particular, while expectancy-value theory focuses on the fact that self-efficacy and value predict student outcomes including their grade, choice of majors and careers, this theory does not focus on how the learning environments improve or deteriorate student self-efficacy and value. We found that the learning environment including peer interaction, perceived recognition, and belonging predicted students' physics identity, self-efficacy and interest at the end of the semester after controlling for their pre-self-efficacy, pre-interest and gender. These findings suggest that an inclusive and equitable learning environment can greatly help improve students' motivational beliefs in physics especially for students who already had lower motivational beliefs because of societal biases and stereotypes about physics even before starting their physics course. In the future studies, we intend to investigate the motivational beliefs of students from different ethnic/racial groups to further study these issues.

### ACKNOWLEDGEMENT
This work was supported by NSF award DUE-1524575.

[1] H. B. Carlone and A. Johnson, Understanding the science experiences of successful women of color: Science identity as an analytic lens, J. Res. Sci. Teach. **44**, 1187 (2007).
[2] Z. Hazari, R. H. Tai, and P. M. Sadler, Gender differences in introductory university physics performance: The influence of high school physics preparation and affective factors, Sci. Educ. **91**, 847 (2007).
[3] Z. Hazari, G. Sonnert, P. M. Sadler, and M.-C. Shanahan, Connecting high school physics experiences, outcome expectations, physics identity, and physics career choice: A gender study, J. Res. Sci. Teach. **47**, 978 (2010).
[4] V. Sawtelle, E. Brewe, and L. H. Kramer, Exploring the relationship between self‐efficacy and retention in introductory physics, J. Res. Sci. Teach. **49**, 1096 (2012).
[5] J. M. Nissen and J. T. Shemwell, Gender, experience, and self-efficacy in introductory physics, Phys. Rev. Phys. Educ. Res. **12**, 020105 (2016).
[6] E. M. Marshman, Z. Y. Kalender, C. Schunn, T. Nokes-Malach, and C. Singh, A longitudinal analysis of students' motivational characteristics in introductory physics courses: Gender differences, Can. J. Phys. **96**, 391 (2017).
[7] E. M. Marshman, Z. Y. Kalender, T. Nokes-Malach, C. Schunn, and C. Singh, Female students with A's have similar physics self-efficacy as male students with C's in introductory courses: A cause for alarm?, Phys. Rev. Phys. Educ. Res. **14**, 020123 (2018).
[8] C. M. Steele and J. Aronson, Stereotype threat and the intellectual test performance of African Americans, J. Pers. Soc. Psychol. **69**, 797 (1995).
[9] G. C. Marchand and G. Taasoobshirazi, Stereotype threat and women's performance in physics, Int. J. Sci. Educ. **35**, 3050 (2013).
[10] N. I. Karim, A. Maries, and C. Singh, Do evidence-based active-engagement courses reduce the gender gap in introductory physics?, Eur. J. Phys. **39**, 025701 (2018).
[11] Z. Y. Kalender, E. Marshman, C. D. Schunn, T. J. Nokes-Malach, and C. Singh, Gendered patterns in the construction of physics identity from motivational factors, Phys. Rev. Phys. Educ. Res. **15**, 020119 (2019).
[12] D. Doucette, R. Clark, and C. Singh, Hermione and the Secretary: how gendered task division in introductory physics labs can disrupt equitable learning, Eur. J. Phys. **41**, 035702 (2020).
[13] E. Lichtenberger and C. George-Jackson, Predicting high school students' interest in majoring in a STEM field: Insight into high school students' postsecondary plans, J. Career Tech. Educ. **28**, 19 (2013).
[14] J. S. Eccles, Understanding women's educational and occupational choices: Applying the Eccles et al. model of achievement-related choices, Psychol. Women Quarterly **18**, 585 (1994).
[15] B. J. Zimmerman, Self-efficacy: An essential motive to learn, Contemp. Educ. Psychol. **25**, 82 (2000).
[16] H. M. Watt, The role of motivation in gendered educational and occupational trajectories related to maths, Educ. Res. Eval. **12**, 305 (2006).
[17] A. L. Zeldin, S. L. Britner, and F. Pajares, A comparative study of the self‐efficacy beliefs of successful men and women in mathematics, science, and technology careers, J. Res. Sci. Teach. **45**, 1036 (2008).
[18] D. H. Schunk and F. Pajares, The development of academic self-efficacy, in *Development of Achievement Motivation: A Volume in the Educational Psychology Series*, edited by A. Wigfield and J. S. Eccles (Academic Press, San Diego, 2002), p. 15.
[19] Z. Y. Kalender, E. Marshman, C. D. Schunn, T. J. Nokes-Malach, and C. Singh, Damage caused by women's lower self-efficacy on physics learning, Phys. Rev. Phys. Educ. Res. **16**, 010118 (2020).
[20] E. Seymour, Tracking the processes of change in US undergraduate education in science, mathematics, engineering, and technology, Sci. Educ. **86**, 79 (2002).
[21] S. G. Brainard and L. Carlin, A six‐year longitudinal study of undergraduate women in engineering and science, J. Engin. Educ. **87**, 369 (1998).
[22] S. J. Correll, Gender and the career choice process: The role of biased self-assessments, Am. J. Sociol. **106**, 1691 (2001).
[23] S. J. Correll, Constraints into preferences: Gender, status, and emerging career aspirations, Am. Sociol. Rev. **69**, 93 (2004).
[24] C. Hill, C. Corbett, and A. St Rose, *Why so Few? Women in Science, Technology, Engineering, and Mathematics* (ERIC, 2010).
[25] A. B. Diekman, E. K. Clark, A. M. Johnston, E. R. Brown, and M. Steinberg, Malleability in communal goals and beliefs influences attraction to STEM careers: Evidence for a goal congruity perspective, J. Pers. Soc. Psychol. **101**, 902 (2011).
[26] A. Maries, N. I. Karim, and C. Singh, Is agreeing with a gender stereotype correlated with the performance of female students in introductory physics?, Phys. Rev. Phys. Educ. Res. **14**, 020119 (2018).
[27] A. Bandura, Self-efficacy, in *Encyclopedia of Psychology*, 2nd ed., edited by R. J. Corsini (Wiley, New York, 1994), Vol. 3, p. 368.
[28] A. Bandura, Social cognitive theory of self-regulation, Organ. Behav. Hum. Decis. Process. **50**, 248 (1991).
[29] S. L. Britner and F. Pajares, Sources of science self-efficacy beliefs of middle school students, J. Res. Sci. Teach. **43**, 485 (2006).
[30] J. M. Harackiewicz, K. E. Barron, J. M. Tauer, and A. J. Elliot, Predicting success in college: A longitudinal study of achievement goals and ability measures as predictors of interest and performance from freshman year through graduation, J. Educ. Psychol. **94**, 562 (2002).
[31] S. Hidi, Interest: A unique motivational variable, Educ. Res. Rev. **1**, 69 (2006).

[32] A. Wigfield and J. S. Eccles, The development of achievement task values: A theoretical analysis, Dev. Rev. **12**, 265 (1992).
[33] A. Wigfield and J. S. Eccles, Expectancy–value theory of achievement motivation, Contemp. Educ. Psychol. **25**, 68 (2000).
[34] Z. Hazari, G. Potvin, R. M. Lock, F. Lung, G. Sonnert, and P. M. Sadler, Factors that affect the physical science career interest of female students: Testing five common hypotheses, Phys. Rev. ST Phys. Educ. Res. **9**, 020115 (2013).
[35] Z. Hazari, P. M. Sadler, and G. Sonnert, The science identity of college students: Exploring the intersection of gender, race, and ethnicity, J. Coll. Sci. Teach. **42**, 82 (2013).
[36] C. Monsalve, Z. Hazari, D. McPadden, G. Sonnert, and P. M. Sadler, Examining the relationship between career outcome expectations and physics identity, in *Proceedings of the Physics Education Research Conference*, Sacramento, CA (2016), p. 228.
[37] G. Potvin and Z. Hazari, The development and measurement of identity across the physical sciences, in *Proceedings of the 2013 Physics Education Research Conference*, Portland, OR (2013), p. 281.
[38] A. Godwin, G. Potvin, Z. Hazari, and R. Lock, Identity, critical agency, and engineering: An affective model for predicting engineering as a career choice, J. Engin. Educ. **105**, 312 (2016).
[39] R. M. Lock, Z. Hazari, and G. Potvin, Physics career intentions: The effect of physics identity, math identity, and gender, AIP Conf. Proc. **1513**, 262 (2013).
[40] Z. Hazari and C. Cass, Towards meaningful physics recognition: What does this recognition actually look like?, The Phys. Teach. **56**, 442 (2018).
[41] Z. Hazari, E. Brewe, R. M. Goertzen, and T. Hodapp, The importance of high school physics teachers for female students' physics identity and persistence, The Phys. Teach. **55**, 96 (2017).
[42] K. A. Renninger and S. Hidi, Revisiting the conceptualization, measurement, and generation of interest, Educ. Psychol. **46**, 168 (2011).
[43] S. Hidi and K. A. Renninger, The four-phase model of interest development, Educ. Psychol. **41**, 111 (2006).
[44] Z. Y. Kalender, E. Marshman, C. D. Schunn, T. J. Nokes-Malach, and C. Singh, Why female science, technology, engineering, and mathematics majors do not identify with physics: They do not think others see them that way, Phys. Rev. Phys. Educ. Res. **15**, 020148 (2019).
[45] M. Meeuwisse, S. E. Severiens, and M. P. Born, Learning environment, interaction, sense of belonging and study success in ethnically diverse student groups, Res. High. Educ. **51**, 528 (2010).
[46] M. Lorenzo, C. H. Crouch, and E. Mazur, Reducing the gender gap in the physics classroom, Am. J. Phys. **74**, 118 (2006).
[47] R. Masika and J. Jones, Building student belonging and engagement: Insights into higher education students' experiences of participating and learning together, Teach. High. Educ. **21**, 138 (2016).
[48] C. Goodenow, Classroom belonging among early adolescent students: Relationships to motivation and achievement, J. Early Adolesc. **13**, 21 (1993).
[49] N. M. Hewitt and E. Seymour, A long, discouraging climb, ASEE Prism **1**, 24 (1992).
[50] A. W. Astin, *What Matters in College: Four Critical Years Revisited* (Jossey-Bass, San Francisco, 1993).
[51] E. Seymour, N. M. Hewitt, and C. M. Friend, *Talking about Leaving: Why Undergraduates Leave the Sciences* (Westview Press, Boulder, CO, 1997), Vol. 12.
[52] D. S. Yeager and G. M. Walton, Social-psychological interventions in education: They're not magic, Rev. Educ. Res. **81**, 267 (2011).
[53] Activation Lab Tools: measures and data collection instruments (2017), http://www.activationlab.org/tools/.
[54] J. Schell and B. Lukoff, Peer instruction self-efficacy instrument [Developed at Harvard University] (unpublished).
[55] S. M. Glynn, P. Brickman, N. Armstrong, and G. Taasoobshirazi, Science motivation questionnaire II: Validation with science majors and nonscience majors, J. Res. Sci. Teach. **48**, 1159 (2011).
[56] PERTS Academic Mindsets Assessment (2020), https://survey.perts.net/share/dlmooc.
[57] L. J. Cronbach, Coefficient alpha and the internal structure of tests, Psychometrika **16**, 297 (1951).
[58] R. Likert, A technique for the measurement of attitudes, Arch. Sci. Psychol. **140**, 5 (1932).
[59] W. S. Gosset, The probable error of a mean, Biometrika **6**, 1 (1908).
[60] J. Cohen, *Statistical Power Analysis for the Behavioral Sciences* (L. Erlbaum Associates, Hillsdale, N.J., 1988).
[61] A. J. Tomarken and N. G. Waller, Structural equation modeling: Strengths, limitations, and misconceptions, Annu. Rev. Clin. **1**, 31 (2005).
[62] D. Hooper, J. Coughlan, and M. Mullen, Structural equation modeling: Guidelines for determining model fit, Electronic J. Bus. Res. Methods **6**, 53 (2007).
[63] K. L. Tonso, Student engineers and engineer identity: Campus engineer identities as figured world, Cult. Stud. Sci. Educ. **1**, 273 (2006).
[64] J. E. Stets and P. J. Burke, A sociological approach to self and identity, in *Handbook of Self and Identity* (The Guilford Press, New York, 2003), p. 128.
[65] Z. Hazari, C. Cass, and C. Beattie, Obscuring power structures in the physics classroom: Linking teacher positioning, student engagement, and physics identity development, J. Res. Sci. Teach. **52**, 735 (2015).